\documentclass[twocolumn,reprint,aps,prd,nofootinbib,floatfix,superscriptaddress]{revtex4-1}
 \usepackage[unicode]{hyperref} 
\usepackage[utf8]{inputenc}
\usepackage{amsmath}
\usepackage{amsfonts}
\usepackage{amssymb}
\usepackage[left=2cm,right=2cm,top=2cm,bottom=2cm]{geometry}
\usepackage{braket}
\usepackage{dsfont}
\usepackage{hyperref}

\usepackage{graphicx}
\usepackage{tikz}
\usetikzlibrary{arrows,decorations.pathmorphing,backgrounds,positioning,fit,petri,shapes}
\usepackage{lipsum}
\usepackage{xcolor}
\usepackage{makecell}
\usepackage{xurl}
\usepackage{booktabs}

\usepackage{multirow}
\begin{document}

\title {Connecting the Dots: Prospects for Detecting Multi-Source Inflation through Scale-Dependent and Stochastic Bias of Little Red Dots}

\begin{abstract}
Recent work has proposed using the spatial clustering of Little Red Dots (LRDs) as a high-redshift cosmological probe. In this paper, we investigate the feasibility of obtaining a LRD sample with a future space experiment, and forecast LRD constraints on primordial local-type non-Gaussianity, $f^{\rm loc}_{\rm NL}$, and the collapsed-limit trispectrum amplitude, $\tau^{\rm loc}_{\rm NL}$. LRDs are ideally suited to probe these inflationary parameters because they span a large cosmological volume, have high bias and are easily identifiable. We perform idealised joint $f_{\rm NL}^{\rm loc}$--$\tau_{\rm NL}^{\rm loc}$ Fisher forecasts assuming the universality relation, Gaussian covariance, Poisson shot-noise and a normalised redshift dispersion, $\sigma(z) / \left( {1+z} \right)= 0.03$. We quantify the significance at which one would detect multi-source inflation under the Suyama–Yamaguchi relation, $\tau^{\rm loc}_{\rm NL} > \left( 6/5 f_{\rm NL}\right)^2$, across the $f_{\rm NL}^{\rm loc}$--$\tau_{\rm NL}^{\rm loc}$ plane. We find that a LRD analysis could robustly detect multi-source inflation at high-significance over a large region of parameter space, and would be more constraining than all other past or present experiments.

\end{abstract}

\author{Peter L.~Taylor}
\email{taylor.4264@osu.edu}
\affiliation{Center for Cosmology and AstroParticle Physics (CCAPP), The Ohio State University, Columbus, OH
43210, USA}
\affiliation{Department of Physics, The Ohio State University, Columbus, OH 43210, USA}
\affiliation{Department of Astronomy, The Ohio State University, Columbus, OH 43210, USA}
\author{Xinyi Chen}
\affiliation{Center for Cosmology and AstroParticle Physics (CCAPP), The Ohio State University, Columbus, OH
43210, USA}
\affiliation{Department of Physics, The Ohio State University, Columbus, OH 43210, USA}
\affiliation{Department of Astronomy, The Ohio State University, Columbus, OH 43210, USA}
\author{Christopher M. Hirata}
\affiliation{Center for Cosmology and AstroParticle Physics (CCAPP), The Ohio State University, Columbus, OH
43210, USA}
\affiliation{Department of Physics, The Ohio State University, Columbus, OH 43210, USA}
\affiliation{Department of Astronomy, The Ohio State University, Columbus, OH 43210, USA}
\author{Charuhas Shiveshwarkar}
\affiliation{Center for Cosmology and AstroParticle Physics (CCAPP), The Ohio State University, Columbus, OH
43210, USA}
\affiliation{Department of Physics, The Ohio State University, Columbus, OH 43210, USA}

\maketitle

\section{Introduction} \label{sec:intro}
Measuring the amplitudes of specific templates of the $N$-point functions of the primordial curvature perturbation field, $\zeta$, may yield insights into the physics of inflation. 
\par Most work to date has focused on the amplitude of the local-type squeezed bispectrum, $f^{\rm loc}_{\rm NL}$. In single-field inflation, there is usually no primordial non-Gaussianity in the squeeze limit~\cite{2004JCAP...10..006C}. Meanwhile multi-field models generally predict $\left| {f_{\rm NL}^{\rm loc}} \right|> 1 $~\cite{2003PhRvD..67b3503L,2003PhRvD..67b3503L,2004PhRvD..69d3508Z, 2017PhRvD..95l3507D}.  \footnote{It should be noted that that not all multi-field models imply $\left| {f_{\rm NL}^{\rm loc}} \right|> 1 $~\cite{2018JCAP...04..027R, 2011PhRvD..83j3517M}, and not all single-field models imply $\left| {f_{\rm NL}^{\rm loc}} \right|<1 $~\cite{2013EL....10139001N, 2013EL....10259001C}}
\par This is not the only informative template. In this paper, we will also consider the amplitude of the collapsed-limit trispectrum amplitude, $\tau^{\rm loc}_{\rm NL}$, originally defined in~\cite{2008PhRvD..77b3505S}. 
\par Informally, $\tau^{\rm loc}_{\rm NL}$ is the amplitude of Fourier-space quadrilaterals that have been “squeezed” along the diagonal. If multiple inflationary fields are present, the value of $\tau^{\rm loc}_{\rm NL}$ determines the number of independent combinations that source primordial perturbations.
\par Formally, under most physical assumptions~\cite{2008PhRvD..77b3505S, 2011PhRvL.107s1301S, 2012NuPhB.864..492K}, $f^{\rm loc}_{\rm NL}$ and $\tau^{\rm loc}_{\rm NL}$ are related by the Suyama–Yamaguchi relation
\begin{equation} \label{eqn:rel}
\tau^{\rm loc}_{\rm NL}\geq
\left(\frac{6}{5}f^{\rm loc}_{\rm NL}\right)^2.
\end{equation}
This result was originally derived in~\cite{2008PhRvD..77b3505S}. Strict inequality in Eqn.~\ref{eqn:rel} implies that multiple independent field combinations source perturbations in $\zeta$, while strict equality implies that only one independent field combination sources perturbations in $\zeta$~\cite{2008PhRvD..77b3505S, 2011PhRvL.107s1301S, 2012NuPhB.864..492K}.   
\par The most precise constraints on these parameters come from Planck measurements of the cosmic microwave background (CMB). Current measurements place $f_{\rm NL}^{\rm loc} = -0.9 \pm 5.1$~\cite{2020A&A...641A...9P}, and $\tau_{\rm NL} ^{\rm loc} < 1500$ at the $95 \%$ confidence level~\cite{2025PhRvD.111l3534P}. However, primary CMB measurements of these parameters are fundamentally cosmic variance limited -- future improvements will be driven by measurements of post-recombination probes.
\par A key prediction of local-type primordial non-Gaussianity, is that it leads to a $1/k^2$-scale dependent halo bias on large scales~\cite{2008PhRvD..77l3514D, 2008JCAP...08..031S}. Compared to other manifestations of primordial non-Gaussianity, this makes $f_{\rm NL} ^{\rm loc}$ much easier to disentangle from non-Gaussianities induced by nonlinear structure growth.
\par Many studies have leveraged this phenomenon to constrain $f_{\rm NL}^{\rm loc}$ e.g., ~\cite{2021arXiv210613725M, 2024MNRAS.532.1902R, 2026JCAP...02..056F}. Using a combination of Luminous Red Galaxies (LRGs) and Quasars (QSOs), constraints from today's leading spectroscopic survey, the Dark Energy Spectroscopic Instrument (DESI), are approaching the precision of Planck: $f_{\rm NL}^{\rm loc} = 3.5^{+10.7}_{-7.4}$~\cite{2025JCAP...06..029C}, $f_{\rm NL}^{\rm loc} = -3^{+12}_{-12}$\cite{2026arXiv260624651B} and $f_{\rm NL}^{\rm loc} = 0.1^{+7.4}_{-7.4}$~\cite{2026PhRvD.113f3552C}. 
\par A new generation of proposed and existing experiments have the potential to drive $\sigma (f_{\rm NL} ^{\rm loc})$ below unity. One of the primary mission objectives of The Spectro-Photometer for the History of the Universe, Epoch of Reionization and Ices Explorer (SPHEREx), launched in 2025, is to achieve this target. Idealised Fisher forecasts for the mission place $\sigma (f_{\rm NL} ^{\rm loc}) =0.2$, for a joint power spectrum and bispectrum analysis~\cite{2014arXiv1412.4872D}. The Vera Rubin Observatory's Legacy Survey of Space and Time also has the potential to constrain $\sigma(f_{\rm NL}^{\rm loc}) < 1$~\cite{2018PhRvD..97l3540S}, but this is heavily reliant on accurate photometric redshifts.
\par Other experiments that are proposed or under construction, including: DESI-II~\cite{2022arXiv220903585S}, Spec-S5~\cite{2026arXiv260807927P}, SIRMOS~\cite{2026arXiv260116761B}, the Wide-field Spectroscopic Instrument~\cite{2024arXiv240512518B} and the Square Kilometre Array~\cite{2020PASA...37....7S} -- all have the potential to place tight constraints on $f_{\rm NL}^{\rm loc}$.
\par While substantial energy has been devoted to $f_{\rm NL} ^{\rm loc}$, measuring $\tau_{\rm NL}^{\rm loc}$ from clustering data is much less mature. The collapsed-limit trispectrum amplitude exhibits a stochastic bias in the galaxy power spectrum~\cite{1,2,3,4}, but we are not aware of any studies which use this to constrain $\tau_{\rm NL}^{\rm loc}$ from data. The primary limitation is that the signal goes as $1/k^4$, requiring enormous cosmological volumes for precise measurements. 
\par {\it In this paper we propose using little red dots as a tracer population to probe an unprecedentedly large cosmological volume in order to place the leading constraints on multi-source inflation}. This work is inspired by~\cite{2026arXiv260906926Z} (hereafter Zebrowski26) which first proposed using LRDs to measure baryonic acoustic oscillations. Such a measurement would be a unique probe of cosmic expansion in the matter-dominated era between $4 < z <9$. Subsequently~\cite{2026arXiv260926553K} proposed using LRDs to constraint $f_{\rm NL}^{\rm loc}$, $f_{\rm NL}^{\rm equil}$ and $f_{\rm NL}^{\rm orth}$, immediately prior to the submission of this paper.
\par Following the discovery of LRDs by the James Webb Space Telescope (JWST)~\cite{2024ApJ...963..129M}, $\mathcal{O}(10^3)$ LRDs have been discovered. Determining the underlying physical mechanism responsible for LRDs remains an active area of research~\cite{a,b,c,d,e,f,g,h,i,j,k,l,m,n,o,p,q,r,s,t,u,2026arXiv260912049W}. 
\par Regardless of the underlying physics, LRDs are an ideal cosmological tracer population with which to constrain the primordial parameters. LRDs cover a large cosmological volume and have high bias -- and as Zebrowski26 notes -- LRDs are easily identifiable from a trifecta of distinctive features. Specifically, they are extremely compact, have a steeply rising red-continuum, and they have an extremely luminous broad H$\alpha$ emission line. 
\par To avoid thermal backgrounds from the telescope and sky absorption, detecting the H$\alpha$ line in the range $4 < z <9$, requires space-based data. The JWST field of view is too small to acquire a sample of LRDs over a large area, so Zebrowski26 proposes a future space mission. This would first identify LRDs photometrically, before switching to spectroscopy to determine redshifts.
\par The two primary objectives of this work are to determine:
\begin{enumerate}
\item the feasibility of obtaining a LRD sample from space; and
\item the detection significance of multi-source inflation across parameter space under the Suyama-Yamaguchi relation. 
\end{enumerate}
{\it The main results are summarised in Eqn.~\ref{eqn:result} and Fig~\ref{fig:results}.}
\par The structure of the paper is as follows. In Sec.~\ref{sec:obs} we investigate the feasibility of constructing a LRD sample with a future space mission. In Sec.~\ref{sec:formalism} we review the power spectrum and Fisher formalisms which are used to forecast the constraining power of the LRD sample in Sec~\ref{sec:results}. We assume a flat $\Lambda$CDM Planck 2018 cosmology~\cite{2020A&A...641A...6P} throughout, unless explicitly state otherwise.
\par While it is not the primary focus of this work, the reader should bear in mind that   measuring $f_{\rm NL}^{\rm loc}$ and $\tau_{\rm NL}^{\rm loc}$ requires control over a large number of theoretical and observational systematics to avoid bias, including: relativistic and wide-angle effects~\cite{2020MNRAS.499.2598W, 2012PhRvD..85d1301B, 2014JCAP...09..037B, 2025JCAP...07..063G}, imaging systematics~\cite{2013PASP..125..705P, 2021MNRAS.506.3439R, 2026ApJ..1000...56H}, the $f_{\rm NL}^{\rm loc} - b_{\phi}$ degeneracy~\cite{2008JCAP...08..031S, 2026PhRvD.113j3523S, 2026arXiv260204987P}, UV-background fluctuations~\cite{2023PhRvD.108j3538S}, redshift uncertainties and non-Poissonian shot-noise~\cite{1, 2015PhRvD..91d3506F, 4}. Care would be needed to fully mitigate these systematics in a real measurement.

\section{Constructing a LRD Sample with a Low-Spectroscopic Resolution Survey from Space}  \label{sec:obs}

\begin{figure*}[!hbt]
    \centering
    \includegraphics[width=\linewidth]{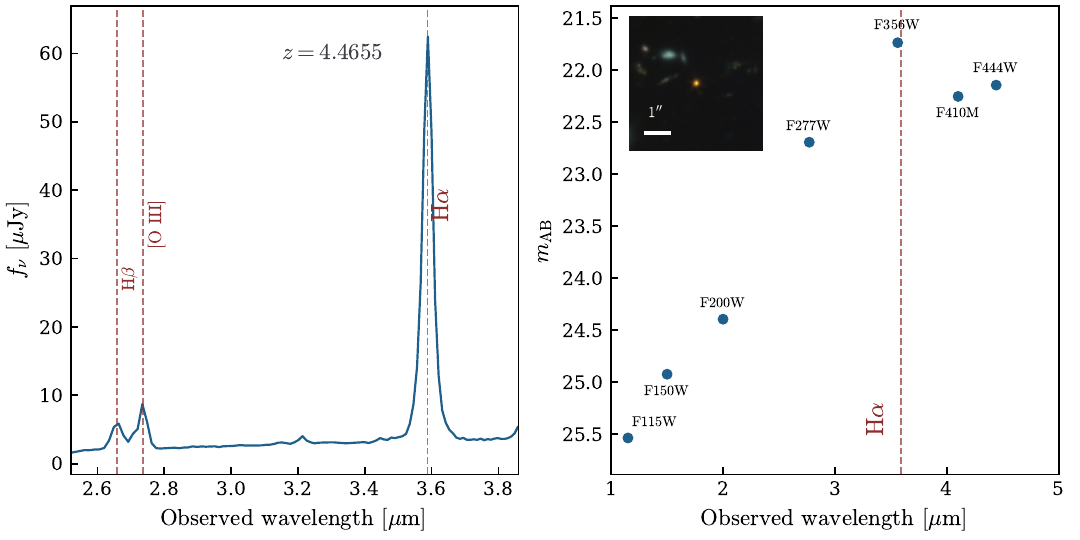}
    \caption{Luminous Little Red Dot A2744-45924, discovered in~\cite{2024arXiv241204557L}.
    {\bf Left:} Spectrum of A2744-45924 featuring a broad luminous H$\alpha$
    emission line.
    {\bf Right:} Photometry exhibiting the characteristic steeply rising red continuum and stacked (F115W, F277W, F444W) cutout of
    A2744-45924.}
    \label{fig:LRD}
\end{figure*}

\subsection{Little Red Dots}
\par LRDs are easily identifiable from their compact morphology, steeply rising red continuum, and luminous and broad H$\alpha$ emission line. This is summarised in Fig~\ref{fig:LRD}.
\par Spectral data\footnote{\url{https://s3.amazonaws.com/msaexp-nirspec/extractions/uncover-v4/uncover-v4_prism-clear_2561_45924.spec.fits}} is obtained using public DJA/UNCOVER DR4 NIRSpec/PRISM data~\cite{2025ApJ...982...51P}. Photometric data is obtained from the UNCOVER DR3 Photometric Catalog~\cite{2024ApJ...976..101S, 2024ApJS..270....7W, 2024ApJ...974...92B}\footnote{ https://doi.org/10.5281/zenodo.11059273}.
Cutouts are generated using the DAWN JWST Archive (DJA)/Grizli cutout API\footnote{\url{https://dawn-cph.github.io/dja/general/api_summary/} and
\url{https://zenodo.org/records/7767790}}.
\par Most LRDs exhibit a characteristic `V-shape' continuum. Targeting this feature may improve selection, but for this paper this paper, we have decided to focus on the rapidly rising red continuum as this feature is found in all LRDs. See~\cite{s} for a potential physical explanation of this phenomenon. 

\subsection{Instrument, Approximations and Survey Design Choices}
\par Leveraging these observational characteristics, we now describe the observational requirements needed to build a LRD sample for primordial non-Gaussianity cosmology. {\it The primary objective of this section is to demonstrate feasibility.}

\par Zebrowski26 proposed a $14,000\  {\rm deg}^2$ space-based survey (leaving room to optimally cut around the galactic and ecliptic planes) using a single broad band to identify candidates, before switching to a single low-resolution spectrograph to obtain redshifts. We adopt the same strategy, but elect to image multiple bands to identify the steeply rising red-continuum.
\par As a baseline, we assume a $D = 2 \ m$ mirror with a $\Omega = 2 \ {\rm deg}^2$ FoV with no 
detector gaps and no instrumental noise.
\par We do not advocate for specific hardware choices, but in this paper, we will consider thirty two $4096 \times 4096$ pixel detectors spread across eight modules. For comparison the Nancy Grace Roman Space Telescope has fifteen HgCdTe H4RG infrared detectors at this pixel resolution~\cite{2024SPIE13092E..0SS}. The eight modules can operate in imaging or spectroscopic mode and the low-resolution spectroscopic systems with $R=\lambda/\Delta \lambda =30$ prisms or grisms are fed by digital micro-mirror devices (DMDs). Dividing the FoV into modules makes the optics more tractable by allowing separate optical relays to correct and demagnify much smaller fields independently. 
\par As a baseline, we envisage a annular-field three-mirror anastigmat (TMA) telescope with an effective focal length, $ f=22 \ m$, so that the f-number is $f/D = 11$. This is similar to the  SuperNova/Acceleration Probe (SNAP) design~\cite{2002SPIE.4849..215L}.
\par The $2 \ {\rm deg}^2$ FoV is then divided into $0.25 \ {\rm deg}^2$ patches across $n_{\rm mod} = 8$ modules that can operate in photometric or spectroscopic mode.  Within each module, relay optics demagnify the light to have a $f$-number, $f/4.69$, corresponding to an effective focal length $f_{\rm eff}=9.38\ {\rm m}$. The effective focal area across the detectors is $A_{\rm foc} = f_{\rm eff}^2 \Omega \simeq 536 \ cm^2$, which matches the area of the 32 H4RG detectors. Assuming a pixel pitch (pixel size) of $p = 10 \ \mu m$~\cite{2020JATIS...6d6001M}, implies an angular pixel scale, $\theta_{\rm pix} = p/f_{\rm eff} \simeq 0.22^{\prime \prime}$.

\par Existing H2RG $2048 \times 2048$ detectors on JWST currently cover the wavelength range of the H$\alpha$ line out to $z\sim7$, while the $1024 \times 1024$ MIRI detectors cover the range $5 < \lambda < 28 \mu m$ and can operate down to $0.6\ \mu m$~\cite{2023PASP..135d8002R}. Using multiple detectors would lead to a more complicated instrument design. 
\par Meanwhile, the Spectroscopic All-sky explorer~\cite{2009ExA....23...39C}, SIRMOS~\cite{2026arXiv260116761B}, and ATLAS~\cite{2019PASA...36...15W} mission concepts have all proposed to use DMDs and this technology has undergone extensive space-environment testing~\cite{2017JATIS...3c5003T}. However, we are not aware of any DMDs that are commercially available for our precise wavelength range. 
\par DMD number count is a potential issue. Our proposed design requires one hundred thirty $2048 \times 1080$  $4.14 \ cm^2$ DMDs~\cite{2009ExA....23...39C} to cover the focal area (or seventeen DMDs per module). One could reduce this number by separating the spectroscopic and photometric optics in each module, and using a combination of coarser detectors and relay mirrors to shorten the effective focal length on the spectroscopic arm. One could also abandon the DMD design altogether and algorithmically separate spectral traces from full-field slitless spectroscopy as in Euclid and Roman.

\subsection{Observational Characteristics of the LRD Source Population}

\par We take the H$\alpha$ equivalent width to be $E_{\rm obs} = 1 \ \mu m$~\cite{2026arXiv260317667M, 2026OJAp....962505S} and the LRD AB limiting magnitude to be $m_{\rm F444W} = 27.5$ at $4.44 \ \mu m$. This is slightly more conservative than the limiting magnitude used in existing samples~\cite{2026ApJ...997...48B, 2025ApJ...991...37A}.  
\par We treat LRDs as point sources, and for source extraction we take an aperture corresponding to the first minimum of the Airy disc $\theta_{\rm ap} = 1.22 \lambda / D$. The LRD energy enclosed within this aperture is $f_{\rm ee} = 0.84$, where we have only accounted for the diffraction from the primary mirror, ignoring other sources of diffraction in the optical system. 

\subsection{The Photometric Survey}

\par We envisage a three band imaging survey. This is summarised in Tab~\ref{tab:imaging_bands}. Again, we make no claim that this is the optimal strategy; the purpose here is to demonstrate feasibility.

\begin{table*}
    \centering
    \caption{Broad-band imaging survey overview. All three bands lie within the wavelength range of existing HgCdTe devices.}
    \label{tab:imaging_bands}
    \setlength{\tabcolsep}{10pt}
    \begin{tabular}{lcccc}
    \hline \hline
    Band & Top-hat range [$\mu$m] &
    $B_\nu$ [MJy sr$^{-1}$] &
    $t_{\rm pointing}$ [hr] & $t_{\rm survey}$ [yr] \\
    \hline
    Short-wave & 1.5--2.5 & 0.12 & 0.182 & 0.146 \\
    Mid-wave   & 2.5--3.5 & 0.15 & 0.753 & 0.601 \\
    Long-wave  & 3.5--5.0 & 0.30 & 2.810 & 2.244 \\
    \hline \hline

    \end{tabular}
\end{table*}

\par To select LRDs on colour from their steeply rising red continuum, we divide the imaging survey into three $1.5  < \lambda < 5. \ \mu  m$ bands (see Tab.~\ref{tab:imaging_bands} for details). For simplicity we adopt an AB limiting magnitude $m_{\rm AB}=27.5$ across all bands\footnote{This is probably optimistic in the bluest band, but this helps identify LRDs in the lowest redshift bin which contains the least cosmological information.}~\cite{2026ApJ...997...48B, 2025ApJ...991...37A,2024ApJ...968...38K} and require a signal-to-noise $S/N =5$, across all bands. As we will show, the survey time is dominated by the long-wavelength band, leaving the possibility of subdividing the bluer bands. One could use additional ground-based observations to identify the Lyman-break. 
\par For selection, we imagine using a compactness criterion, only selecting sources with PSF-dominated light profiles~\cite{b}. This requires resolving LRDs to near the resolution of the PSF. The pixel side-length is $0.22^{\prime\prime}$, while the na\"ive PSF size from the diffraction limit of the primary mirror is $0.21^{\prime\prime}$ at $2 \ \mu m$. We will assume this resolution is sufficient for the morphological selection, which is complemented by the other selection criteria.
\par Assuming good thermal control and an absence of stray light contamination, zodiacal light is the dominant background~\cite{2023PASP..135d8002R} over the range $1.5  < \lambda < 6.65 \ \mu  m$. We take the zodiacal background surface brightness to be $B_\nu = (0.12,\,0.15,\,0.30)\,\ \mathrm{MJy\,sr^{-1}}$ for the three bands in order of increasing wavelength. This is in line with measurements from JWST~\cite{2023PASP..135d8002R}, although we stress that the zodiacal background can vary by $\sim 50 \%$ depending on the time of year and location of the target~\cite{2023PASP..135d8002R}.

\subsection{The Spectroscopic Survey}

We consider two possible wavelength ranges: case 1: $3.3 < \lambda < 6.65 \ \mu m$ (covering the H$\alpha$ line for $4<z<9$) and  case 2: $3.3 < \lambda < 5.25 \ \mu m$ (covering the H$\alpha$ line for $4<z<7$) and mandate a signal-to-noise of $S/N = 5$ for the H$\alpha$ line detection. The survey specifications for case 1 and case 2 are summarised in Tab~\ref{tab:lrd_surveys}. For the second case, one could use existing H4RG detectors. 

\begin{table}
    \centering
    \small
    \setlength{\tabcolsep}{4pt}
    \caption{Summary of the two LRD spectroscopic survey options.
    The zodiacal background $B_\nu$ is evaluated at the maximum redshift
    of each survey. Tomographic bins bin boundaries lie at $z=(4,5,6,7,9)$.}
    \label{tab:lrd_surveys}
    \begin{tabular}{lcc}
        \hline \hline
        & {\bf case 1} & {\bf case 2} \\
        \hline
        Redshift range
            & $4<z<9$
            & $4<z<7$ \\

        Wavelength range [$\mu{\rm m}$]
            & $3.3$--$6.65$
            & $3.3$--$5.25$ \\

        Tomographic bin count
            & $4$
            & $3$ \\

        $B_\nu$ [$\mathrm{MJy\,sr^{-1}}$]
            & $2.0$
            & $0.6$ \\

        Pointing time [yr]
            & $1.1$
            & $0.9$ \\
        \hline
        \hline
    \end{tabular}
\end{table}

\par  The zodiacal background rapidly increases with wavelength~\cite{2023PASP..135d8002R}, so to be conservative we assume that all sources lie at $z=9$ ($\lambda = 6.65 \ \mu  m$) for case 1 and $z=7$ ($\lambda \approx 5.25 \ \mu  m$) for case 2. At these wavelength we set the zodiacal background surface brightness to be $B_\nu = 2  \ {\rm M J} y \ {\rm sr}^{-1}$ and $B_\nu = 0.6  \ {\rm M J} y \ {\rm sr}^{-1}$, in line with current JWST observations (see Fig. 1 in ~\cite{2023PASP..135d8002R}). 
\par We will assume high spectroscopic targeting purity. Since the LRDs are sparse on the sky, and given the high spatial resolution and low spectroscopic resolution of the detectors, we will assume that there are no spectral trace collisions.

\subsection{Photometric Survey Time Calculation}
Assuming Poisson noise, the signal-to-noise is given by the ratio of the source photon count to the square root of the total photon count
\begin{equation} \label{eq:1}
\frac{S}{N}
=
\frac{\dot{N}_{\rm src}\,t}
{\sqrt{
\left(
\dot{N}_{\rm src}
+
\dot{N}_{\rm zod}
\right)t
}},
\end{equation}
where $\dot{N}_{\rm zod}$ is the zodiacal background photon count, $\dot{N}_{\rm src}$ is the source photon count, and $t$ is the exposure time. Further assuming each band is a top hat with constant flux density, $f_{\nu}$, across the band we find
\begin{equation}
\dot N_{\rm src}
 = Af_{\rm ee} \int_{\nu_{\rm min}}^{\nu_{\rm max}}\frac{f_\nu}{h\nu}\,d\nu
 = \frac{A f_\nu f_{\rm ee}}{h}\ln\!\left(\frac{\lambda_{\rm max}}{\lambda_{\rm min}}\right),
\end{equation}
where $A = \pi (D/2)^2$ is the area of the mirror, $\nu$ is the frequency, $D$ is the diameter of the primary, $\lambda_{\rm max /min}$ are the band-boundaries and
\begin{equation}
f_\nu
=
3631\,{\rm Jy}\times10^{-0.4 m_{\rm AB}}.
\end{equation}
Similarly, the zodiacal background is
\begin{equation}
\dot N_{\rm zod}
=\frac{A\Omega_{\rm ap}B_\nu}{h}
 \ln\!\left(\frac{\lambda_{\rm max}}{\lambda_{\rm min}}\right),
\end{equation}
where $\Omega_{\rm ap}$ is the solid angle of the extraction aperture, $\Omega_{\rm ap} = \pi \theta_{\rm ap}^2$. Solving Eqn.~\ref{eq:1} for $t$ yields pointing times from which one can readily calculate the survey time. This yields pointing times of $3.7$ hours per field, for total imaging time of $3.0$ years. A detailed breakdown of the per-band imaging times is given in Tab~\ref{tab:imaging_bands}. 

\subsection{Spectroscopic Survey Time Calculation}
\par With these choices, we now calculate the pointing time, $t$. For the purposes of this calculation, we further assume a rest optical continuum follows a $T_{\rm eff} = 4050 \ K$ blackbody spectrum with surface brightness $B_\nu (\lambda^{\rm em}, T_{\rm eff})$~\cite{2026OJAp....962505S}.
\par Assuming Poisson noise, the signal-to-noise is given by the ratio of H$\alpha$ photon count to the square root of the photon count from all sources, so that
\begin{equation} \label{eqn:s/n}
\frac{S}{N}
=
\frac{\dot{N}_{\rm line}\,t}
{\sqrt{
\left(
\dot{N}_{\rm line}
+\dot{N}_{\rm cont}
+\dot{N}_{\rm zod}
\right)t
}}.
\end{equation}
Here the H$\alpha$ photon line rate is
\begin{equation}
\dot N_{\rm line} = \frac{AF_{\rm line} f_{\rm ee} }{h \nu},
\end{equation}
where $\nu$ is the observed frequency of the H$\alpha$ line, and the H$\alpha$ line flux is,
\begin{equation}
F_{\rm line}
=
f_\lambda(\lambda_{\rm H\alpha}^{\rm obs})\,E_{\rm obs},
\end{equation}
where $f_\lambda(\lambda^{\rm obs}_{H \alpha})$ is the flux density of the continuum at the wavelength of the $H \alpha$ line in the observed frame. It is given by
\begin{equation}
f_\lambda^{\rm obs}(\lambda_{\rm H\alpha}^{\rm obs})
=
\frac{c}{(\lambda_{\rm H\alpha}^{\rm obs})^2}
f_\nu^{\rm obs}(\lambda_{\rm H\alpha}^{\rm obs}).
\end{equation}
where
\begin{equation}
f_\nu^{\rm obs}(\lambda_{\rm H\alpha}^{\rm obs})
=
f_\nu^{\rm obs}(\lambda_{\rm cont}^{\rm obs})
\frac{
B_\nu(\lambda_{\rm H\alpha}^{\rm em},T)
}{
B_\nu(\lambda_{\rm cont}^{\rm em},T)
},
\end{equation}
and $f_\nu^{\rm obs}(\lambda_{\rm cont}^{\rm obs})$ is flux density of the continuum at $4.44 \ \mu m$ and is given by,
\begin{equation}
f_\nu^{\rm obs}(\lambda_{\rm cont}^{\rm obs})
=
3631\,{\rm Jy}\times10^{-0.4m_{\rm F444W}},
\end{equation}
$B_\nu(\lambda_{\rm H\alpha}^{\rm em},T)$ is the continuum surface brightness in the rest frame at the wavelength of the H$\alpha$ emission line and $B_\nu(\lambda_{\rm cont}^{\rm em},T)$ is the continuum surface brightness at $4.44 \ \mu m$.

Meanwhile the continuum photon rate is
\begin{equation}
\dot N_{\rm cont} = \frac{Af_{\nu} \Delta \nu f_{\rm ee}}{h \nu},
\end{equation}
 and the photon rate of the zodiacal background is,
\begin{equation}
\dot N_{\rm zod} = \frac{ A B_\nu \Delta \nu \Omega_{\rm ap}}{h \nu},
\end{equation}
Solving Eqn.~\ref{eqn:s/n} for $t$ in case 1 and case 2 yields pointing times of $1.3$ and $1.1$ hours respectively. This yields survey times of  $1.1 \ {\rm years}$ and $0.9 \ {\rm years}$.

\subsection{Total Survey Time}
In summary, the total exposure times for case 1 and case 2 assuming $100 \%$ throughput are
\begin{equation} 
    t_{\rm exposures} \simeq
    \begin{cases}
        3.9~{\rm yr}, & \text{case 1: } 4<z<7, \\[3pt]
        4.1~{\rm yr}, & \text{case 2: } 4<z<9.
    \end{cases}
\end{equation}
Imaging in the long-wavelength band consumes approximately half of this time. If we further assume $60 \%$ throughput and $20 \%$ overhead time, the total survey time is
\begin{equation} \label{eqn:result}
    \boxed{
    t_{\rm survey} \simeq
    \begin{cases}
        8.1~{\rm yr}, & \text{case 1: } 4<z<7, \\[3pt]
        8.5~{\rm yr}, & \text{case 2: } 4<z<9.
    \end{cases}
    }
\end{equation}
{\it Power spectrum constraints on the inflationary parameters, particularly $\tau_{\rm NL}^{\rm loc}$ are primarily sample-variance limited, so it might be possible to reduce the survey time or instrumental requirements if we do not require the full sample down to $m_{\rm AB} = 27.5$. Covering a larger area could also improve constraints.} A detailed analysis is left to future work.

\subsection{Redshift Precision}
For the cosmological analysis, we need to account for the redshift uncertainties of the sources. By using the full line profile, one can typically centroid an emission line to higher precision than the width of a spectral element. However, primordial non-Gaussianity is sensitive to large scales, so we do not require high redshift precision. Instead we note the redshift dispersion is $\sigma(z) / ({1+z})= \sigma(\lambda) / \lambda$, and on taking the wavelength dispersion to be the width of a spectral element, $\sigma(\lambda) = \Delta \lambda$, we arrive at a conservative estimate of the normalised redshift dispersion\footnote{This approximation is permissible because the full width half maximum of a typical LRD H$\alpha$ emission line is $1500 \ {\rm km\,s^{-1}}$,~\cite{2025A&A...701A.168D} is an order of magnitude smaller than the full width half maximum of the instrument, $c/R \approx 10,000 \ {\rm km\,s^{-1}}.$},
\begin{equation}
\frac{\sigma(z)}{1+z}  = \frac{1}{R} \approx0.03.
\end{equation}
We will use this value throughout the remainder of the paper.

\section{Theory} \label{sec:formalism}

\subsection{Fisher Formalism}
We closely follow the formalism outlined in~\cite{2014arXiv1412.4872D} and restrict our attention to the power spectrum.

\begin{table*}
    \centering
    \caption{Fiducial Little Red Dot tracer properties used in this work.
    The number densities, linear biases, and survey volumes are adopted from
    Ref.~\cite{2026arXiv260906926Z}. The fundamental mode is defined as
    $k_f=2\pi/V_{i}^{1/3}$. The choice of cubic geometry is chosen to keep consistency with the SPHEREx forecasts.}
    \label{tab:lrd_tracers}
    \begin{tabular}{cccccccc}
        \hline\hline
        $z$ range
        & $z_{\rm center}$
        & $\bar n\,[h^3\,{\rm Mpc}^{-3}]$
        & $V_{\rm i}\,[h^{-3}{\rm Gpc}^3]$
        & $k_f\,[h\,{\rm Mpc}^{-1}]$
        & $b_1$
        & $nP(0.01 \ h {\rm Mpc}^{-1})$
        & $nP(0.2\ h {\rm Mpc}^{-1} )$ \\
        \hline
        $4.0$--$5.0$ & 4.5 & $6.3\times10^{-5}$ & 46.70
        & $1.74\times10^{-3}$ & 3.4 & 0.717 & 0.073 \\

        $5.0$--$6.0$ & 5.5 & $1.3\times10^{-4}$ & 41.75
        & $1.81\times10^{-3}$ & 4.5  & 1.859 & 0.190 \\

        $6.0$--$7.0$ & 6.5 & $1.6\times10^{-4}$ & 37.35
        & $1.88\times10^{-3}$ & 5.8  & 2.858 & 0.292 \\

        $7.0$--$9.0$ & 8.0 & $1.1\times10^{-4}$ & 63.82
        & $1.57\times10^{-3}$ & 8.1  & 2.664 & 0.272 \\
        \hline\hline
    \end{tabular}
\end{table*}

\begin{table*}
    \centering
    \caption{Marginalised constraints on local primordial non-Gaussianity.
    The three-bin forecasts omit the highest-redshift bin of the four-bin
    analysis.}
\label{tab:png_constraints}
\setlength{\tabcolsep}{7pt}
\begin{tabular*}{\textwidth}{@{\extracolsep{\fill}}cccccc@{}}
\hline\hline
Analysis
& $N_z$
& $(f_{\rm NL}^{\rm loc})^{\rm fid}$
& $(\tau_{\rm NL}^{\rm loc})^{\rm fid}$
& $\sigma(f_{\rm NL}^{\rm loc})$
& $\sigma(\tau_{\rm NL}^{\rm loc})$ \\
\hline
$f_{\rm NL}$--$\tau_{\rm NL}$ & 4 & 0 & 0 & 0.362 & 2.720  \\
$f_{\rm NL}$--$\tau_{\rm NL}$ & 4 & 1 & 130 & 0.614 & 21.138  \\
$f_{\rm NL}$ only & 4 & 0 & 0 & 0.265 & -- \\
$f_{\rm NL}$--$\tau_{\rm NL}$ & 3 & 0 & 0 & 0.636 & 7.637 \\
$f_{\rm NL}$--$\tau_{\rm NL}$ & 3 & 1 & 130 & 0.922 & 32.920  \\
$f_{\rm NL}$ only & 3 & 0 & 0 & 0.443 & --  \\
\hline\hline
\end{tabular*}
\end{table*}

\begin{figure*}[!hbt]
\includegraphics[width = \linewidth]{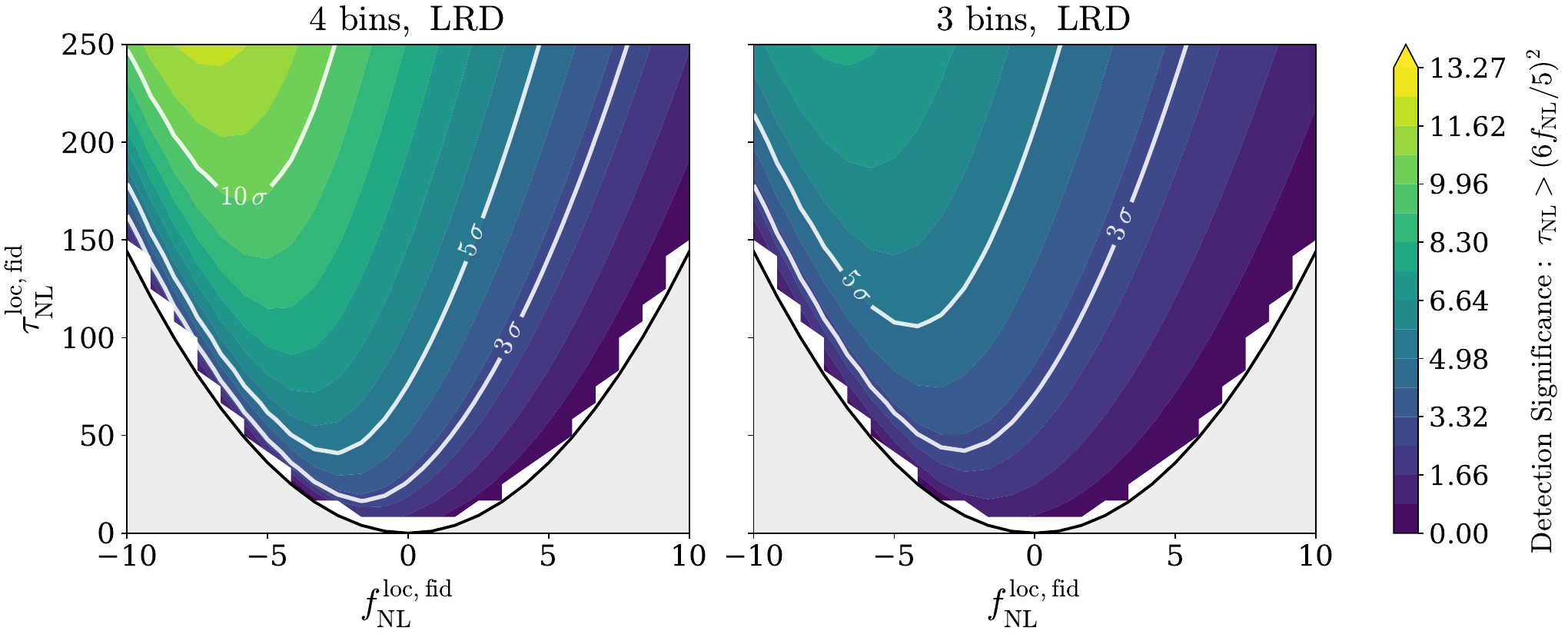}
\caption{Multi-Source Inflation detection significance under the Suyama-Yamaguchi relation (see Sec.~\ref{sec:sig} for details). Fisher detection significances may be overestimated near the $A_\tau =0$ boundary, as we do not impose a physical prior, $\tau^{\rm loc}_{\rm NL} -
\left(\frac{6}{5}f^{\rm loc}_{\rm NL}\right)^2 \geq 0$. {\bf Left:} Fiducial LRD four-bin analysis (case 1). {\bf Right:} LRD three-bin analysis (case 2).}
\label{fig:results}
\end{figure*}

\par We perform a Fisher analysis over the parameters
\begin{equation}
\boldsymbol{\theta}
=
\left(
\omega_b,\omega_c,h,n_s,\sigma_8,
f_{\rm NL}^{\rm loc}, \tau_{\rm NL}^{\rm loc},
b_1,b_2,b_3,b_4
\right),
\end{equation}
where $b_{i}$ is the linear LRD bias in bin $i$. 
The Fisher matrix is given by
\begin{equation} \label{eqn:fisher}
  F^{\rm gg}_{\alpha\beta} =
  \sum_i \frac{V_i}{8\pi^2}
  \int_{k_{f,i}}^{k_{\max}} dk\,k^2
  \int_{-1}^{1} d\mu\,
  \frac{\partial_\alpha P_g\,\partial_\beta P_g}
       {(P_g + 1/\bar n_i)^2},
\end{equation}
where $V_{i}$ is the volume of tomographic bin $i$ (see Tab.~\ref{tab:lrd_tracers}), $k_{f, i} = 2 \pi / V_i^{1/3}$ is the fundamental mode, $k_{\rm max} = 0.2 \ h{\rm Mpc^{-1}}$ throughout, $\bar n_i$ is the galaxy number density in bin $i$, $P_g = P(k,\mu)$ is the power spectrum and $\mu$ is cosine to the line of sight. The choice of the fundamental mode assumes a cubic geometry which is admittedly overly optimistic, but this keeps our forecasts roughly in line with the SPHEREx analyses referenced in this analysis which used $k_{\rm min} = 0.001 \ h{\rm Mpc^{-1}}$. Parameter covariances are found by inverting the Fisher matrix. 
\par We use a two-sided finite difference with a step size of $5 \%$ (and absolute step size of 0.05 for $f_{\rm NL}^{\rm loc}$ and $\tau_{\rm NL}^{\rm loc}$) to compute the derivatives in Eqn.~\ref{eqn:fisher}. We confirm that the results are insensitive to the exact choice of step size to first decimal precision.
\par We take a Planck 2018 prior for the $\Lambda$CDM parameters, so that the combined Fisher matrix is 
\begin{equation}
    F_{\alpha\beta} = F^{ gg }_{\alpha\beta} + F_{\alpha\beta}^{\rm Planck}.
\end{equation}

\subsection{Scale Dependent Bias in the Power Spectrum}
Following~\cite{2014arXiv1412.4872D, 2008PhRvD..77l3514D, 2008JCAP...08..031S, 1, 2015PhRvD..91d3506F, 1}, we write the power spectrum in bin $i$ as
\begin{equation} \label{eq:power}
\begin{split}
P_g(k,\mu,z_i) ={}&
D_z(k,\mu,z_i)
\Big[
\left(b_i + f_i\mu^2
+ f_{\rm NL}^{\rm loc} C_i(k,z_i)\right)^2 \\
&\qquad\qquad
+ A_\tau C_i^2(k,z_i)
\Big] P_m(k,z_i).
\end{split}
\end{equation}
Here $P_m(k,z_i)$ is the linear matter power spectrum computed by ${\tt pycamb}$~\cite{2000ApJ...538..473L}. We include a Kaiser term~\cite{1987MNRAS.227....1K} and redshift uncertainty damping term, 
\begin{equation}
D_z(k,\mu,z)
= \exp\left[-(k\mu\sigma_r)^2\right],
\end{equation}
to account for the redshift uncertainty~\cite{2014arXiv1412.4872D}. Here $\sigma_r=c\sigma(z) / H(z)$, where $H(z)$ is the Hubble rate. Meanwhile we define,
\begin{equation}
A_\tau
\equiv
\frac{25}{36}
\left[
\tau_{\rm NL}^{\rm loc}
-
\left(
\frac{6}{5}f_{\rm NL}^{\rm loc}
\right)^2
\right].
\end{equation}
and the scale-dependent bias coefficient is,
\begin{equation}
C_i(k,z)
=
\frac{3\,\Omega_m(H_0/c)^2}{2\,k^2T(k)D(a)}
\,b_{\phi,i}
\end{equation}
 where $\Omega_m$ is the matter-density parameter, $D(a)$ is the linear growth normalised to the scale factor, $a$, in the matter dominated era as in~\cite{2014arXiv1412.4872D,2024JCAP...05..094S}, $T(k)$ is the linear transfer function and $b_{\phi,i}$ is the response of the LRD bias in bin $i$ to primordial perturbations. 
 \par We further assume the universality relation 
 \begin{equation}
b_{\phi,i} = 2\delta_c\left(b_i - 1\right),
 \end{equation}
 where $\delta_c \approx 1.686$ is the critical linear overdensity for spherical collapse. This is a good approximation for halos that have not undergone recent mergers~\cite{2008JCAP...08..031S}. In the future, one would need to carefully construct a prior on $b_{\phi,i}$ from simulations. For other tracer populations, one typically finds a $\sim 15 \%$ uncertainty on $b_{\phi,i}$~\cite{2024PhRvD.109j3530H,2022JCAP...01..033B,2020JCAP...12..013B,2017MNRAS.468.3277B}. 
 \par Finally we assume Poissonian stochasticity, since non-Poisson shot-noise is degenerate  with $\tau_{\rm NL}^{\rm loc}$~\cite{1, 2015PhRvD..91d3506F, 4}. To account for this in the future, one could marginalise over a free stochastic bias term~\cite{2015PhRvD..91d3506F}, or calibrate the stochasticity with simulations over a large volume~\cite{2013PhRvD..88h3507B}. 
 
 \subsection{Parameter Dependence of the Data Covariance}
In $\Lambda$CDM, the underlying parameters are well determined, so the data covariance is fairly insensitive to the true underlying cosmology~\cite{2019OJAp....2E...3K, 2009A&A...502..721E, 2017MNRAS.465.4016R}. For example, in the absence of shot-noise, the covariance of the power spectrum on large scales scales as $\sigma_8^4$, so choosing $\sigma_8 = 0.83$, rather than  $\sigma_8 = 0.8$ when estimating the covariance, only leads to a $\sim 15 \%$ change in the value of covariance matrix elements.
\par This is not the case for scale-dependent bias measurements of $f_{\rm NL}^{\rm loc}$ and $\tau_{\rm NL}^{\rm loc}$~\cite{2024JCAP...05..094S}. To see this, refer to Eqn.~\ref{eq:power}, and note that in the absence of shot-noise, the data covariance scales as $P_{\rm g} ^2$ and the galaxy power spectrum itself scales as $\left( f_{\rm NL}^{\rm loc} \right)^2$ and $\tau_{\rm NL}^{\rm loc}$. Since the uncertainties on the magnitude of $ f_{\rm NL}^{\rm loc} $ and $\tau_{\rm NL}^{\rm loc}$ are currently of $\mathcal{O}(1)$ and $\mathcal{O}(10^3)$, the covariance will vary strongly across the choice of fiducial cosmology. For this reason, we will investigate the parameter dependence of our results across the $f_{\rm NL}^{\rm loc}$--$\tau_{\rm NL}^{\rm loc}$ plane in detail.

 \subsection{Multi-Source Inflation Detection Significance} \label{sec:sig}
 We now show how to compute the detection significance of multi-source inflation for a point in parameter space centered at $\left( f_{\rm NL}^{\rm loc, fid}, \tau_{\rm NL}^{\rm loc, fid} \right)$, using the Suyama–Yamaguchi relation. We define
 \begin{equation}
  D \equiv \tau_{\rm NL} - \left(\frac{6 f^{\rm loc}_{\rm NL}}{5}\right)^2,
\end{equation}
so that the detection significance, $S$, is
\begin{equation}
  S = \frac{D}{\sigma(D)},
\end{equation}
where the error, $\sigma(D)$, is found by expanding $D$ to first order and using the Fisher formalism,
 \begin{equation}
    \sigma_D^2 =
    \sum_{\alpha,\beta \in \{f^{\rm loc}_{\rm NL},\,\tau_{\rm NL}^{\rm loc}\}}
    \partial_\alpha D\,
    \left(F^{-1}\right)_{\alpha\beta}\,
    \partial_\beta D .
  \end{equation}

\section{Results} \label{sec:results}
\subsection{Baseline Fisher Results}
We perform Fisher calculations under of a variety of analysis choices. The results are summarised in Tab~\ref{tab:png_constraints}.
\par For our baseline analysis, we assume 4 tomographic bins (case 1), and fiducial values $f_{\rm NL}^{\rm loc, fid} = \tau_{\rm NL}^{\rm loc, fid} = 0$.\footnote{This choice of fiducial value is for illustrative purposes only. The Fisher formalism breaks down near the boundary $A_\tau = 0$, if one were to impose the physical prior, $A_\tau \geq 0$.} The marginalised errors on the primordial parameters are $\sigma(f^{\rm loc}_{NL}) = 0.4$ and $\sigma(\tau^{\rm loc}_{NL}) = 2.7$. 
\par Next, we consider the three tomographic bin survey (case 2), which drops LRDs above $z = 7$. For fiducial values $f_{\rm NL}^{\rm loc, fid} = \tau_{\rm NL}^{\rm loc, fid} = 0$, we find $\sigma(f_{\rm NL}^{\rm loc}) = 0.6$ and $\sigma(\tau_{\rm NL}^{\rm loc}) = 7.6$.  Dropping the fourth bin may reduce the survey time and simplify the instrument design, but this choice nearly triples the uncertainties on $\tau_{\rm NL}^{\rm loc}$ in our baseline analysis.

\subsection{Parameter Sensitivity}
To test the response to fiducial parameter choice, we repeat the $4$-bin analysis centred on $f_{\rm NL}^{\rm loc, fid} = 1$ and $\tau_{\rm NL}^{\rm loc, fid} = 130$. In this case, $\sigma(f_{\rm NL}^{\rm loc}) = 0.6$ and $\sigma(\tau_{\rm NL}^{\rm loc}) = 21.1$. Comparing to the baseline case, $f_{\rm NL}^{\rm loc, fid} = \tau_{\rm NL}^{\rm loc, fid} = 0$, the errors on ${\tau_{\rm NL}^{\rm loc}}$ have increased by a factor of six. In general, constrains degrade with increasing $|f_{\rm NL}^{\rm loc, fid}|$ or $\tau_{\rm NL}^{\rm loc, fid}$.

\subsection{Comparison with Other Surveys}
Our baseline $4$-bin survey (case 1) yields $\sigma(f_{\rm NL}^{\rm loc}) = 0.36$. This is more constraining than the current Planck constraints $\sigma(f_{\rm NL}^{\rm loc}) = 5.1$, forecasted Spec-S5 constraints, $\sigma(f_{\rm NL}^{\rm loc}) = 1.4$,~\cite{2025arXiv250307923B} and forecasted SPHEREx power spectrum constraints $\sigma(f_{\rm NL}^{\rm loc}) = 0.87$, but it is less constraining than the joint power spectrum/bispectrum SPHEREx forecasts, $\sigma(f_{\rm NL}^{\rm loc}) = 0.2$. 
\par However, the real power of the LRD analysis comes when one wishes to jointly constrain $f_{\rm NL}^{\rm loc}$ and $\tau_{\rm NL}^{\rm loc}$~\cite{2014arXiv1412.4872D}. At $f_{\rm NL}^{\rm loc, fid} = \tau_{\rm NL}^{\rm loc, fid} = 0$, $\sigma(\tau_{\rm NL}^{\rm loc}) = 2.7$. This is orders of magnitude smaller than the leading Planck trispectrum constraint, $\tau_{\rm NL} ^{\rm loc} < 1500$.
\par Meanwhile at $f_{\rm NL}^{\rm loc, fid} = 1$ and $\tau_{\rm NL}^{\rm loc, fid} = 130$ we  find $\sigma(\tau_{\rm NL}^{\rm loc}) = 21.1$. This is a substantial improvement over the anticipated SPHEREx power spectrum constraint, centred on the same value in the $f_{\rm NL}^{\rm loc}$ -- $\tau_{\rm NL}^{\rm loc}$ plane, $\sigma(\tau_{\rm NL}^{\rm loc}) = 78$~\cite{2024JCAP...05..094S}.

\subsection{Multi-Source Inflation Detection Significance}
In this subsection, we investigate the significance with which one could detect signatures of multi-source inflation using the Suyama–Yamaguchi relation (Eqn.~\ref{eqn:rel}) across $\left( f_{\rm NL}^{\rm loc}, \tau_{\rm NL}^{\rm loc} \right)$ space.
\par To do this we compute the detection significance following the formalism of Sec.~\ref{sec:sig} on a $25 \times 31$ grid over 
$f_{\rm NL}^{\rm loc} \in [-10,10]$ and $\tau_{\rm NL}^{\rm loc} \in [0,250]$. We do this for the three and four redshift bin cases. The results are summarised in Fig.~\ref{fig:results}.
\par We find that in all cases, one would be able to detect signatures of multi-source inflation at the $3\sigma$ and $5\sigma$-levels over a substantial fraction of the space. 
\par We have not explicitly impose a physical Suyama-Yamaguchi prior, $A_\tau \geq 0$ and this would reduce the detection significance near the boundary if we were to perform a full MCMC analysis~\cite{2024JCAP...05..094S}. 
\par The inclusion of the highest-redshift LRD bin helps substantially. Interestingly, detection is stronger for negative values of $f_{\rm NL}^{\rm loc, fid}$. This is because the power spectrum (Eqn.~\ref{eq:power}) contains terms that are even and odd powers of $f_{\rm NL}^{\rm loc}$.

\section{Conclusion} \label{sec:conclusion}

We investigated the possibility of using the scale-dependent bias of a Little Red Dot (LRD) tracer population to constrain multi-source inflation, using the Suyama-Yamaguchi relation. 
\par LRDs are ideally suited for this purpose as they trace a large cosmological volume between $4<z<9$, and are easily identifiable from their compact morphology, steeply rising red continuum, and luminous H$\alpha$ emission line.
\par No telescope is currently capable of constructing such a sample, so we investigate the feasibility of obtaining this data with a future observatory. To avoid thermal backgrounds, this can only be achieved from space -- although ground based data to identify e.g., the Lyman-break feature may be complimentary.
\par In an idealised scenario, a two-meter telescope with a $2 \ \rm deg^2$ FoV could obtain the required imaging and spectroscopic data across $14,000\  {\rm deg}^2$ in approximately eight years. About half the pointing time is spent imaging the long-wavelength band as the zodiacal background rises sharply with increasing wavelength.
\par The survey estimates presented here should be viewed as an order-of-magnitude assessment of the needs for an LRD survey mission, rather than a concrete proposal. Nevertheless, some of the technology needs are already clear: (i) large mosaics of infrared detectors, with modestly more area than Euclid and Roman and with longer-wavelength sensitivity; (ii) cooling of a much larger focal plane than on previous passively cooled missions with 5 $\mu$m cutoff detectors; and (iii) mosaics of DMDs or other slit-forming technology, including buttability and cold qualification. The packaging of the optical path for such a mission will also present a challenge, given the large number of components and the small space if one is to compress the beam to $\sim f/4.5$ to minimise detector and DMD area. One of the first tasks would be to examine the trade space of optical solutions (e.g., surveys in series, versus parallel with specialized channels; in the latter case, split fields versus dichroics) to assess the feasibility of the various options.
\par These headlines results also assume a limiting magnitude $m_{\rm AB} = 27.5$. As the relevant scales for our scientific objectives are dominated by sample-variance -- not shot-noise -- this may be too deep. {\it Targeting a sub-sample of bright LRDs would substantially reduce the cost of the mission, with little loss in constraining power.} Increasing the survey area would also improve constraints, and is worth investigating in a future study.
\par Next, we reviewed the Fisher formalism and the theory of scale dependent and stochastic bias of the galaxy power spectrum -- carefully outlining all simplifying assumptions in the analysis. We argued that the Fisher results are highly sensitive to the choice of fiducial cosmology, as the data covariance is sensitive to the underlying value of $f_{\rm NL}^{\rm loc}$ and $\tau_{\rm NL}^{\rm loc}$. 
\par We used the Fisher formalism to forecast LRD constraints. We showed that LRD measurements of $f_{\rm NL}^{\rm loc}$ would be competitive with today's leading surveys, before demonstrating that LRDs would dramatically sharpen joint $\left( f_{\rm NL}^{\rm loc}, \tau_{\rm NL}^{\rm loc} \right)$ constraints.
\par Finally, we computed the detection significance of multi-source inflation across the $f_{\rm NL}^{\rm loc}$ -- $\tau_{\rm NL}^{\rm loc}$ plane using the Suyama–Yamaguchi relation. We found that it would be possible to detect signatures of multi-source inflation at $>5 \sigma$ across a large area of allowable $f_{\rm NL}^{\rm loc}$ -- $\tau_{\rm NL}^{\rm loc}$-parameter space. 
\par Probing primordial physics with LRDs shows promise, but future work will be needed to define instrumental requirements and determine the full impact of observational and theoretical systematics. Improvements to the analysis could come from:  the inclusion of bispectrum information~\cite{2026arXiv260926553K,2009ApJ...703.1230J}, splitting the LRD sample by luminosity (as a proxy for bias) to reduce cosmic variance through sample variance cancellations~\cite{2009PhRvL.102b1302S, 2009JCAP...10..007M} and cross-correlating with CMB lensing could help mitigate systematics~\cite{2014MNRAS.441L..16G,2025A&A...698A.177B}. Space missions must follow a narrow trajectory through the Overton window -- but if any of today's surveys robustly detect signatures of primordial non-Gaussianity, $| f_{\rm NL}^{\rm loc}| >1$ -- it would be timely to connect the dots.

\section{Acknowledgments}
PLT thanks Ashley Ross, Klaus Honscheid, Josephine Baggen and participants of The Ohio State University's Astro Coffee. This work was supported in part by a grant of access to OpenAI models through the ChatGPT for Academic Researchers program. PLT thanks Jung-Tsung Li for access to this resource. Codex (GPT-5.5 medium, GPT-5.6 Sol) was used for coding, web search, data retrieval, latexing equations and tables, and literature synthesis. GPT found relevant results that were previously unknown to PLT, but these were all traced back to existing literature. The paper was written entirely by the authors, and responsibility for its contents remain with the authors. Claude (Opus 5.5) was used to independently check the manuscript and rerun the entire analysis after the first draft was assembled as a secondary validation check. PLT and XC acknowledge the support of the U.S. Department of Energy (DOE) award number DE-SC0011726. This research used resources of the National Energy Research Scientific Computing Center (NERSC), a Department of Energy User Facility. CS acknowledges support from the David and Lucile Packard Foundation; NASA grant 22-ROMAN11-0011, via a subaward from the Jet Propulsion Laboratory; and the SPHEREx project under a contract from the NASA/Goddard Space Flight Center to the California Institute of Technology.
\newline
{\it Authorship Contributions:} PLT led the analysis and wrote the paper. XC and CS provided theoretical expertise and guidance on inflationary physics. CMH provided expertise and guidance on survey design and instrumentation. 

 \bibliographystyle{apsrev4-1.bst}
\bibliography{bibtex.bib}

\appendix

\end{document}